\documentclass[12pt]{article}

\begin{document}

\begin{center}
{\huge \bf Towards the explanation of flatness
of galaxies rotation curves} \\[10mm]
 S.A. Larin \\ [3mm]
 Institute for Nuclear Research of the
 Russian Academy of Sciences,   \\
 60-th October Anniversary Prospect 7a,
 Moscow 117312, Russia
\end{center}

\vspace{30mm}
\begin{abstract} 
We suggest a new explanation of flatness of galaxies rotation curves
without invoking dark matter.
For this purpose a new gravitational tensor field is introduced
in addition to the metric tensor.
\end{abstract}

\newpage

\section{Introduction}

Velocities of stars and gas rotation  around galaxies centers
become independent of rotation radii at large enough radii,
see e.g. \cite{mcg} and references therein.
This asymptotical flatness of galaxies rotation curves is in obvious contradiction with
Newtonian dynamics which demands that velocities should decrease with radii
as $1/\sqrt{r}$. 

The contradiction is explained within  the dark matter paradigm \cite{dm}. 
Within the paradigm a galaxy is placed in a spherical halo of dark matter \cite{hal}
with mass density of dark matter decreasing as $1/r^2$. Hence the Newtonian gravitational potential becomes
logarithmically dependent on $r$ providing a flat rotation curve.
But hypothetical constituents of dark matter  are still not discovered in direct experiments
inspite of numerous searches.

In this situation it is worthwhile to develope explanations of flatness 
of galaxies rotation curves which do not invoke dark matter.
Probably the most known attempt of this type is 
modified Newtonian dynamics, called MOND \cite{milg}-\cite{mil2} where it is assumed
that gravitational forces depend on accelerations of objects participating
in interactions. Modified Newtonian dynamics has essential phenomenological
successes. In particular it described the known Tully-Fisher relation \cite{tf}
which establishes  correlation between a galaxy luminosity and a corresponding
 flat rotation speed. But modified Newtonian dynamics  does not describe e.g. 
gravitational lensing by galaxies and galaxies clusters.
Modified Newtonian dynamics is not a completely formalized theory
although there was an attempt to construct its complete Lagrangian version
\cite{be} which was not supported experimentally.

It should be mentioned that the Tully-Fisher relation is more precise
in its barionic form which states that a flat rotation speed in a galaxy
correlates with its barionic mass i.e. a sum of stars and gas masses, see
\cite{mcga} and references therein. The barionic Tully-Fisher relation states that 
$M_{bar} \propto V_{flat}^4$.

This tight correlation between visible matter of a galaxy
and a corresponding flat rotation speed
is also a rather strong motivation to find the explanation
of flatness of galaxies rotation curves without dark matter.
Also in \cite{mc1} it is shown that there is a strong correlation between
the observed radial star acceleration and
the acceleration predicted by the observed distribution of baryons.

It is necessary to point out that, over the years, there have also been other ideas to replace 
the dark universe paradigm. For example through the so 
called de Sitter relativity \cite{t1}-\cite{t4} or through other approaches \cite{te1}-\cite{te3}.

In the present paper
we suggest a new explanation of flatness of galaxies rotation curves
without invoking dark matter.
For this purpose a new tensor gravitational field is introduced
in addition to the metric tensor.

\section{Main part}
We consider the General Relativity action  
plus tems with a new gravitational tensor field $f_{\mu\nu}$:
\begin{equation}
\label{action}
S=\int d^{4}x  \sqrt{-g} (-M_{Pl}^2 R +g_{\mu\nu}T^{\mu\nu}
+M_{Pl}^2 \Lambda
\end{equation}
\[ 
+ f_{\mu\nu}(x)(D^{\lambda}D_{\lambda})^{3/2} f^{\mu\nu}(x)
+G^* \sum_i\frac{1}{\sqrt{1+m_i/m^*}}f_{\mu\nu} T^{\mu\nu}_i),
\]
here the $R$-term is the Einstein-Hilbert Lagrangian of General Relativity,
the geometrical part of the action;
$\sqrt{-g}$  is as usual the square root of the minus determinant of the metric tensor
$g_{\mu\nu}(x)$.
$M_{Pl}^2=1/(16\pi G)$ is the Planck mass squared.

$D_{\lambda}$ is the standard covariant derivative, for example:
\begin{equation}
D_{\lambda}f^{\mu\nu}=\frac{\partial}{\partial x^{\lambda}}f^{\mu\nu}
+\Gamma^{\mu}_{\lambda\sigma}f^{\sigma\nu}
+\Gamma^{\nu}_{\lambda\sigma}f^{\mu\sigma},
\end{equation}
where the Christoffell symbols as usual are
\begin{equation}
\Gamma_{\mu\nu}^{\alpha}=\frac{1}{2}g^{\alpha\beta}\left(\partial_{\nu}g_{\mu\beta}
+\partial_{\mu}g_{\nu\beta}-\partial_{\beta}g_{\mu\nu}\right).
\end{equation}
$f_{\mu\nu}(x)$ is the introduced new tensor field additional to 
the metric tensor $g_{\mu\nu}(x)$. The field $f_{\mu\nu}(x)$ can be chosen to be
the symmetric tensor similar in this sence to the metric tensor $g_{\mu\nu}(x)$. It is necessary to ensure
that the new gravitational field $f_{\mu\nu}(x)$ interacts with light which has traceless energy-momentum
tensor $T^{\mu\nu}_{light}(x)$. Interaction of the field $f_{\mu\nu}$ with light
is important in order to produce extra gravitational lensing due to barionic matter
as compared to General Relativity, since in the case of absence of dark matter
General Relativity alone is not sufficient to describe observed gravitational lensing.
In principle, the new dynamical gravitational field $f_{\mu\nu}$ can also have
asymmetric part in addition to the symmetric one, but this is not essential
at the present stage of considerations.

The main new point of the action in eq.(\ref{action}) is the non-integer power $3/2$
of the factor $D^{\lambda}D_{\lambda}$. This provides the $1/(k^2)^{3/2}$ behavior of the propagator
of the new dynamical gravitational field $f_{\mu\nu}$ in the momentum space, where $k^2=k_{\mu}k^{\mu}$
is the square of the  four momentum $k_{\mu}$ . Due to such a behavior of the propagator one can 
generate the logarithmic with the distance $r$ gravitational $log(r)$ potential, as it will be
shown below,
which allows to describe flatness of galaxies rotation curves without invoking dark matter.

$G^*$ is the introduced new coupling constant of interaction of the new gravitatonal tensor 
field $f_{\mu\nu}$ 
with matter fields desribed by energy-momentum tensors $T^{\mu\nu}_i$.
Of course, it it assumed that the complete action also contains terms describing
propogations of matter fields in addition to their interactions with gravitational fields.

$m^*$ is the introduced new mass parameter which is necessary to compensate
the dimensions of masses $m_i$ in the corresponding square rootxs. 

The sum $\sum_i$ in the action (\ref{action}) goes over matter objects drscribed by energy-momentum tensors $T^{\mu\nu}_i$
and having masses $m_i$. Couplings of them with the field $f_{\mu\nu}$ depend
on $m_i$ via $\frac{1}{\sqrt{1+m_i/m^*}}$; this property is quite
essential and is used below to reproduce
the famous barionic Tully-Fisher relation.

Numerical values of new constants $G^*$ and $m^*$ should be fixed from fitting 
experiments. {\bf It needs an analysis of a huge amount
of empirical data for different galaxies and that is why it is a subject for a separate publication. }

The $\Lambda$-term  in eq.(\ref{action}) is not essential in perturbation theory 
which we will consider.

We will use the standard in Quantum Field Theory system  of units $\hbar=c=1$.

To quantize the theory (\ref{action}) selfconsistently one should add to
the Lagrangian all possible terms quadratic in the Riemann tensor
$R_{\mu\nu\rho\sigma}$, see \cite{lar1}, \cite{lar2} where 
perturbatively reormalizable
and unitary model of quantum gravity was for the first time formulated.
But these terms are not essential for the present considerations.


As it was already mentioned above we will work within perturbation theory, hence
a linearized theory  around the flat  metric $\eta_{\mu\nu}$
is considered, that is we make the following substitution
\begin{equation}
g_{\mu\nu}=\eta_{\mu\nu}+h_{\mu\nu},
\end{equation}
here the generally accepted  in Field Theory convention in four dimensions is chosen
$\eta_{\mu\nu}=diag(+1,-1,-1,-1)$. 
Indexes are raised and lowered in the following by means of
the flat metric tensor $\eta_{\mu\nu}$.


Within perturbation theory one makes the standard shift of the metric field
\begin{equation}
 h_{\mu\nu} \rightarrow M_{Pl} h_{\mu\nu}.
\end{equation}
Perturbative expansion goes as usual in the  inverse powers of $M_{Pl}$ 
or in other words in the powers of the  Newton coupling constant $G=\frac{1}{16\pi M_{Pl}^2}$.

Let us now get the propagator of the new gravitational tensor field $f_{\mu\nu}$
in the momentum space. For this purpose
we take the quadratic in the field $ f_{\mu\nu}$
part of the Lagrangian of the action in the equation (\ref{action})
and perform the Fourier transform to the momentum space:
\begin{equation}
\label{forma}
Q=
i (2\pi)^4\int d^4 k~ f^{\mu\nu}(-k)\left[ (k^2)^{3/2}
 \eta_{\mu\rho}\eta_{\nu\sigma}\right]
f^{\rho\sigma}(k),
\end{equation}
 

To obtain the  propagator $D_{\mu\nu\rho\sigma}$ of the field $f_{\mu\nu}$ 
one should in the standard way invert the matrix in the  brackets of the  equation (\ref{forma}):
\begin{equation}
[Q]_{\mu\nu\kappa\lambda}D^{\kappa\lambda\rho\sigma}
=\frac{1}{2}(\delta_{\mu}^{\rho}\delta_{\nu}^{\sigma}
+\delta_{\mu}^{\sigma}\delta_{\nu}^{\rho}).
\end{equation}
Then the propagator has the following form
\begin{equation}
\label{prop}
D_{\mu\nu\rho\sigma}=\frac{1}{2i(2\pi)^4}
\frac{\eta_{\mu\rho}\eta_{\nu\sigma}+\eta_{\mu\sigma}\eta_{\nu\rho}}
{(k^2)^{3/2}}.
\end{equation}

The obtained  propagator (\ref{prop}) generates the  logarithmic in the distance
$r$ gravitational  potential.

To derive the logarithmic potential from the propagator (\ref{prop})
one should consider in the standard way the Fourier transform of the propagator in three space
dimensions: 
\begin{equation}
\label{loga}
D(r) \propto \int d^3 k \frac{ e^{i\vec{k}\vec{r}} }{k^3}=-\frac{4\pi}{r} 
\int_{\kappa}^{\infty}d~k\frac{sin(k~r)}{(k)^2},
\end{equation}
where the expression in the right hand side is obtained after integrations over angles are done. 
$\kappa>0$ is a small regularizing parameter, which regularizes the infrared divergency in the
integral.
$k$ is the lenth of the three vector $\vec{k}$.

Then one gets after performing integration in the right hand side of the equation (\ref{loga}):
\begin{equation}
D(r) \propto -Ci(r\kappa)-\frac{sin(r\kappa)}{r\kappa}=-log(r\kappa)+constant +O(\kappa^2),
\end{equation}
where $Ci(r\kappa)$ is the standard cosine integral function, and  
in the right hand side the expansion in the limit of small $\kappa$ is done.

Thus one obtains the logarithmic in $r$ potential due to the field $f_{\mu\nu}$. 
The regularizing  parameter $\kappa$ is inessential
when one takes the derivative in $r$ in order to produce the gravitational force from the
obtained logarithmic potential.

At large enough distances $r$, that is at galactic and extragalactic scales,
this $\log{r}$- potential  starts to dominate over the Newtonian
$1/r$- potential which is generated by the metric field $h_{\mu\nu}$
of  General Relativity for small
potentials.


Thus the gravitational field $f_{\mu\nu}$ generates a $1/r$- force
between a point object with a mass $M_{bar}$
having the energy-momentum tensor
$T_{\mu\nu}=\delta^0_{\mu}\delta^0_{\nu}M_{bar}\delta^3(x)$
(describing a galaxy with the baarionic mass $M_{bar}$ 
of stars plus gas )
and an analogous object with a mass $M_{star}$
(describing a star with the mass $M_{star}$):
\begin{equation}
\label{force}
F=(G^*)^2 \frac{M_{bar}}{\sqrt{1+M_{bar}/m^*}}
\frac{M_{star}}{\sqrt{1+M_{star}/m^*}} \frac{1}{r}.
\end{equation}
At a galaxy mass $M_{bar}$ large compared to the mass parameter $m^*$ 
the square root $\sqrt{1+M_{bar}/m^*}$ becomes approximately just  $\sqrt{M_{bar}/m^*}$,
and the above relation (\ref{force}) takes the form
\begin{equation}
\label{force1}
F \approx (G^*)^2 \sqrt{M_{bar}\cdot m^*}
\frac{M_{star}}{\sqrt{1+M_{star}/m^*}}\frac{1}{r}.
\end{equation}
From the other side according to the second Newton law one gets
\begin{equation}
\label{force2}
F=M_{star}\frac{V_{flat}^2}{r}.
\end{equation}
Equating the expressions (\ref{force1}) and (\ref{force2}) we obtain the
following relation
\begin{equation}
\label{force3}
 M_{bar} \approx \frac{V_{flat}^4}{(G^*)^4}
\frac{1+M_{star}/m^*}{ m^*}.
\end{equation}
The expression (\ref{force3}) reproduces the barionic Tully-Fisher relation
which states that $M_{bar}\propto  V_{flat}^4$.

It is interesting to note that the right hand side of
the expression (\ref{force3}) has the dependence
on the star mass $M_{star}$. 
 
Thus flat rotation speeds of stars satisfy the barionic Tully-Fisher relation
in our model.

We should also mention once more that the introduced tensor field $f_{\mu\nu}$ interacts
with light and that is why adds additional, as compared to the metric
field $h_{\mu\nu}$ of General Relativity,
gravitational lensing due to barionic matter.

\section{Discussions}
The Newtonian theory of gravity could be assumed to be a perfect
theory at the galactic and extragalactic distances. But velocities
of stars and gas rotation as  it is known from the experimental observations
are usually essentially larger than
velocities generated by visible barionic matter as they are estimated 
according to the Newtonian dynamics. It is presently commonly accepted to explain
this paradox by the presence of the necessary amount of dark matter in galaxies.
Also observed gravitational lensing by galaxies and clusters of galaxies 
is larger then lensng which can be produced
within General Relativity due to visible matter only. This is again
traditionally  explained
by the presence  of the appropriate amount of dark matter.
But constituents of dark matter are still not found in spite of the numerous
experimental efforts. In this situation it is worthwhile to develope models
alternative to General Relativity although it is clear that approximations of these
models for the solar system scales should coincide with General Relativity
which is perfectly experimentally tested in the solar system.

\section{Conclusions}
We suggested a new explanation of flatness of galaxies rotation curves without
invoking dark matter.
For this purpose a new tensor gravitational field is introduced
in addition to the metric tensor.
Flat rotation speeds of stars
in our model satisfy the known barionic Tully-Fisher relation.

\section{Acknowledgments}
The author is grateful to colleagues from the Theory Division of 
the Institute for Nuclear Research
for helpful discussions.

\end{document}